\documentclass[final,3p,times,twocolumn]{elsarticle}

\usepackage{amsmath}
\usepackage{amssymb}
\usepackage{amsthm}

\begin{document}

\begin{frontmatter}


\author{Rahul Nigam\corref{cor1}\fnref{label2}}
\ead{rahul.nigam@hyderabad.bits-pilani.ac.in}

\title{Volume of a rotating black hole in 2+1 dimensions}


\author[BITS Pilani Hyderabad Campus]{Suraj Maurya}

\affiliation[BITS Pilani Hyderabad Campus]{BITS Pilani Hyderabad Campus={Physics Department, BITS Pilani Hyderabad Campus},
            addressline={Jawahar Nagar}, 
            city={Hyderabad},
            postcode={500078}, 
            state={Telangana},
            country={India}}

\author[BITS Pilani Hyderabad Campus]{Sashideep Gutti}
\author[BITS Pilani Hyderabad Campus]{Rahul Nigam}

\begin{abstract}
 In this article we apply the technique for maximal volume estimation of a black hole developed by Christodoulou and Rovelli \cite{CR} for Schwarzchild blackhole and by Zhang et al \cite{BZ2} for non rotating BTZ black hole, to the case of a rotating black hole in 2+1 dimensions. We derive the equation of the maximal hypersurface for the rotating BTZ blackhole using the Lagrangian formulation demonstrated in \cite{CR}. Further we use maximization technique illustrated earlier by Bengtsson et al \cite{BJ} for Kerr black hole to arrive at the similar result for our case. We argue that the maximum contribution to the volume of the hypersurface comes from what we call the steady state radius, which we show depends on mass M and  the AdS length scale. We demonstrate that this steady state radius can be arrived at using independent considerations of vanishing extrinsic curvature. We show that the volume of this segment of the maximal hypersurface, the CR volume, depends on mass, AdS length scale and angular momentum J. We further compute the entropy of a scalar field living on the maximal hypersurface for a near extremal black hole and show that it is proportional to the horizon entropy of the black hole.
\end{abstract}




\end{frontmatter}


\section{Introduction}
While a black hole horizon and its area can be defined independent of the coordinate basis, the 3 dimensional volume of a black hole bounded by the horizon depends on the choice of the basis. Computing the volume involves choosing a space-like hypersurface which is a constant time surface. And different coordinate bases have different time coordinate. A frame or basis dependent volume is not an unfamiliar concept. Even in flat Minkowski space, the volume of a region depends on the frame of observer. There has been lot of work estimating the interior volume of a black hole. This question is important since one can then connect the total volume hidden in the black hole to the degrees of freedom on the horizon. There are few proposals about how to estimate the volume in the interior of a black hole. All such volumes have the same boundary given by the black hole horizon. Parikh \cite{MP} had offered a definition of the black hole volume by an invariant slice of the space-time inside the black hole horizon. Parikh's \cite{MP} work was tested for the case of  2D dilaton black holes by D. Grumiller \cite{DG}. DiNunno et al \cite{DM} elucidated several examples of volume inside Schwarzschild black hole for different time slicings. \cite{FTK}, \cite{CGK}, \cite{GGW} have discussed application of idea of the maximal slicing with the aim of studying the thermodynamical properties of a black hole in a semi-classical approach. Gibbons et al. \cite{CGK} discussed the thermodynamical volume, $V_{th}$ inside a black hole in presence of a varying cosmological constant $\Lambda$. $V_{th}$ is defined as conjugate variable to $\Lambda$ appearing in the first law of thermodynamics for black hole - $dE = TdS + \Omega dJ + \Phi dQ + V_{th} d\Lambda$, where $E$ is the enthalpy of the space time. One may also consider a gauge coupling constant $g$ instead of $\Lambda$ arising as the vacuum expectation value.  \\
Christodoulou and Rovelli \cite{CR} provided a somewhat different definition of the black hole volume, in which the volume grows indefinitely as a function of the advance time. We will call this the CR volume. The method to compute the CR volume was shown to be analogous to finding the geodesic equation for a particle in space-time. They found that the volume increases as $V \sim 3\sqrt{3} \pi m^2 v$, in the limit $v >> m$, where $v$ is the advance time. This work was extended to Kerr black holes by \cite{BJ} and to RN black holes by \cite{OYC}. An interesting extension of this work was presented by Yen \cite{YCO}, who computed the CR volume for a certain class of topological black holes and showed that the CR volume is not always a monotonically increasing function of the horizon area.  
The asymptotic estimation of the volume of a blackhole where rotation is involved is accomplished in \cite{BJ} for Kerr blackhole and by \cite{Chew} for Kerr AdS blackhole. Our goal is to extend the Lagrangian analysis done in \cite{CR} to the case involving rotations. The case involving rotations is not a straightforward generalization of the Schwarzschild case due to the presence of inner and outer horizons. Complication also arises due to the fact that the axial Killing vector is not hypersurface orthogonal to the spacelike hypersurface between the horizons. We therefore first develop the CR equations for the simpler case of BTZ blackhole with rotation. In this work, we extend the CR analysis to the case of a rotating BTZ black hole and obtain the necessary equations to locate the maximal hypersurface between the outer and inner horiozons.  \\ 
Given the CR volume enclosed by a black hole horizon, one may introduce a quantum field and statistically compute the entropy. This has been been studied by Zhang et al \cite{BZ}, \cite{BZ2}, Bhaumik et al \cite{MA}, \cite{MA2} and Wang et al \cite{XI}. These groups followed different techniques to compute the entropy and arrive at a slightly different answer. However the real takeaway is that the entropy residing in the CR volume is proportional to the horizon entropy. In this work we show that this result can be extended to the case of a rotating BTZ black hole in the near extremal limit. 

\section{Setup}
The metric of a BTZ black hole - 
\begin{equation}
ds^2 = -N^2(r) dt^2 + N^{-2}dr^2 + r^2 (N^\phi dt + d\phi)^2
\end{equation}
where $N^\phi = -J/2r^2$ and the lapse function $N^2 = -\Lambda r^2 - M + J^2/4r^2$. Defining new coordinates $v$ and $\theta$ as - \begin{equation}
v = t + \int^r{\frac{dr'}{N^2(r')}}, 
\theta = \phi - \int^r{\frac{N^\phi(r')}{N^2(r')}dr'}
\end{equation}
The metric can now be written as,
\begin{equation}
ds^2 = -\bigg[-\Lambda r^2 - M\bigg] dv^2 + 2dvdr - Jdvd\theta + r^2 d\theta^2
\label{btzmetric1}
\end{equation}
$v$ is the advance time. We consider a space with $\Lambda = -1/l^2$ which means the cosmological constant is negative and the background space is AdS. We want to find a hypersurface $\Sigma$ which is bounded between the event horizon and the inner radius such that the volume of $\Sigma$ can be made as large as possible. The event horizon and the inner horizon are given by the zeros of the lapse function $N^2$ and the lapse function vanishes at two values of $r$ - 
\begin{equation}
r_{\pm}^2 = Ml^2\bigg[\frac{1}{2} (1 \pm \sqrt{1 - J^2/M^2l^2})\bigg]
\end{equation}
\section{Maximum volume}
Rovelli et al \cite{CR} have shown that for the volume of the maximal space like hyper surface constituting the volume of a black hole, at late times much after the collapse, the maximum contribution comes from a constant value of the radial coordinate $r$. And we can ignore the parts of the hypersurface which have move along the radial direction. With this information, we first investigate the constant $r$ piece of the hyper surface. For a rotating BTZ black hole, the metric is
\begin{equation} \label{metric}
ds^2 = -\bigg(r^2/l^2 - M\bigg) dv^2 - Jdvd\theta + r^2 d\theta^2
\end{equation}    
So the volume for the constant r hypersurface is 
\begin{equation}
V_\Sigma = \int \sqrt{g} dv d \theta = 2\pi v \sqrt{ -(r^2/l^2 -M)r^2 - J^2/4}
\end{equation}
This is the volume of the black hole interior for a given $\Sigma$. We maximize this volume by choosing a value of the radial parameter,$r_{max}$, which is a solution of $dV_\Sigma/dr = 0$. This gives $r_{max} = l\sqrt{M/2}$, which can be written in terms of the horizon radii as, $r_{max} = (r_+ + r_-)/2$. So $r_{max}$ sits right in between the inner and the outer horizon. And plugging the expression for $r_{max}$, we find $V_{\Sigma max} = \pi v \sqrt{M^2l^2 - J^2}$. Note that this volume increases monotonically with the advance time. At late times this part of the hypersurface which corresponds to constant $r$ will contribute to the significant part of the volume and one can ignore the parts of the hypersurface close to the inner and outer  horizons having a non-zero radial component.\\  
The above result can also be verified using another method. We noted that the maximum contribution to the volume is due to the surface $r=const$. It is argued in Zhang \cite{BZ} that extrinsic curvature of such a  surface should vanishes. We now then evaluate the extrinsic curvature of the surface $r = cosnt$. The unit normal to the surface $r=const$ for the metric \ref{btzmetric1} is given by $n^{\alpha}=(1/\sqrt{f}, \sqrt{f},0)$, where $f(r) = r^2/l^2 - M$. The condition for the vanishing extrinsic curvature is given by $n^{\alpha}_{;\alpha}=0$. We note that the volume element of spacetime with metric given by eq. \ref{btzmetric1} is $\sqrt{|g|}=r$. So the condition for the vanishing extrinsic curvature reduces to 
\begin{equation}
\frac{1}{r}\frac{\partial}{\partial r}\left(r \sqrt{r^2/l^2 -  M}\right)=0
\end{equation}
This gives the steady state radius to be $r_{ss}=l\sqrt{\frac{M}{2}}$ which is same as $r_{max}$ we derived above to maximize the volume.

\section{The Lagrangian formulation}
In this section we develop the Lagrangian formulation developed by \cite{CR} for the case of rotating black hole in 3 dimensions. We write the spacelike hypersurface $\Sigma$, where it's topology is  is a direct product of  $S^1$ and $R$. The coordinates on the hypersurface are given by $\theta$ and $\lambda$. We consider a curve  $\gamma$ in the 2-dimensional space $(v,r)$ paramterized by $\lambda$. This $\lambda$ is used as coordinate on $\Sigma$.
\begin{eqnarray}
\Sigma \sim \gamma \times S^1 \\
\gamma \sim (v(\lambda), r(\lambda))
\end{eqnarray} 
The line element of the induced metric on the hypersurface $\Sigma$ is
\begin{equation}
ds^2_{\Sigma} = \big(-f(r)\dot{v}^2 + 2\dot{v}\dot{r} \big)d\lambda^2 - J\dot{v} d\lambda d\theta  + r^2 d\theta^2
\end{equation}
The dot represents derivative with respect to the parameter $\lambda$ and $f(r) = r^2/l^2 -M$. For $\Sigma$ to be a space-like surface, we require
\begin{equation}
\big(r^2(-f(r)\dot{v}^2 + 2\dot{v}\dot{r}) - J^2\dot{v}^2/4\big) > 0
\end{equation}
And the proper volume of $\Sigma$ is given by 
\begin{equation} \label{volume}
V_{\Sigma}[\gamma] = 2 \pi\int_0^{\lambda_f} d\lambda \sqrt{\big(r^2(-f(r)\dot{v}^2 + 2\dot{v}\dot{r}) - J^2\dot{v}^2/4\big)} 
\end{equation}
This volume depends on the curve $\gamma$,  and  we are looking for the curve $\gamma$ which would maximize this volume. This can be viewed as an extremization problem and our goal is to find the equations of motion for the Lagrangian - 
\begin{equation}
L(r,\dot{r},v,\dot{v}) = \sqrt{\big(r^2(-f(r)\dot{v}^2 + 2\dot{v}\dot{r}) - J^2\dot{v}^2/4\big)}
\end{equation}
We will find the equations of motion and then use appropriate normalization to set 
\begin{equation}\label{lagrangian}
L(r,\dot{r},v,\dot{v}) = 1
\end{equation}
This normalization ensures that $\Sigma$ is space-like. 
The curve $\gamma$ parameterized by $\lambda$ has two end points which we choose to be at $r(\lambda_i = 0) = r_+$ and $r(\lambda_f) =r_-$. Similarly $v(0) = v$ and $v(\lambda_f) = v_f$.

\subsection{Equations of motion}


The Lagrangian obtained above for the BTZ case is defined in terms of $\dot{v}$, $\dot{r}$ and $r$. $v$ is a cyclic coordinate and hence we define an integration constant $a$. Now we find the equation of motion for $v$

\begin{equation}
\partial L/\partial \dot{v} = r^2 \frac{(-2f(r)\dot{v} 2\dot{r} -\frac{J^2\dot{v}}{2})}{2L} = a
\end{equation}
We can write this as 
\begin{eqnarray}\label{vdot}
\dot{v} \bigg( -2r^2f(r) - J^2/2 \bigg) = 2a - 2\dot{r} r^2 \\ 
\dot{v} = \frac{\dot{r}r^2 - a}{r^2 f(r) + J^2/4} \nonumber
\end{eqnarray}

Now using eq.\ref{lagrangian} and eq.\ref{vdot}, we get
\begin{equation}
\dot{v} = \frac{1}{a + \dot{r}r^2}    
\end{equation}

And plugging this expression in eq. \ref{lagrangian}, we find
\begin{equation}\label{rdot}
\dot{r} = - \frac{\sqrt{r^2f(r) + J^2/4 + a^2}}{r^2}    
\end{equation}
The expression for $\dot{v}$ is same as deduced by others for the case of BTZ black hole. And the expression for $\dot{r}$ reduces to the expression for BTZ black hole as shown by \cite{BZ} when we put $J = 0$. 

\subsection{The steady state radius and volume}
As we discussed earlier, most of the contribution to the volume of the maximal hypersurface comes from a steady state phase which corresponds to constant $r_{ss}$. We get the expression for $r_{ss}$ by taking $\dot{r} = 0$ and equating the right hand side of eq. \ref{rdot} to zero. 
\begin{eqnarray}
r_{ss}^2(r_{ss}^2/l^2 - M) + J^2/4 + a^2 = 0 \\ 
\Rightarrow r_{ss}^2/l^2 = \frac{M}{2} \pm \frac{\sqrt{M^2 - \frac{4(J^2/4 + a^2)}{l^2}}}{2}
\end{eqnarray}
We showed earlier in section 3 that the steady state radius $r_{ss} = l\sqrt{\frac{M}{2}}$. Demanding that the two methods should give the same value for $r_{ss}$, we can write 
\begin{equation}
    M^2 - \frac{4(J^2/4 + a^2)}{l^2} = 0
\end{equation}
This gives us the value for the constant of motion $a$ as
\begin{equation}\label{a}
    a = \sqrt{\frac{M^2l^2 - J^2}{4}}
\end{equation}
Note that for $\dot{r} = 0$, we have  $\dot{v} = 1/a$. As $v$ must be a monotonically increasing function of the parameter $\lambda$, $a$ must be positive definite and that is why we have taken the positive root. 
\\
Now the volume of the hypersurface as defined in eq. \ref{volume}, while $L(r,\dot{r},v,\dot{v}) =1$, can be evaluated to be 
\begin{equation}
    V_{\Sigma} = 2\pi \int d\lambda = 2\pi a \int dv = 2\pi a v
\end{equation}
Substituting the expression for $a$ as in eq. \ref{a}, we find 
\begin{equation}
V_{\Sigma} = \pi v \sqrt{M^2l^2 - J^2}
\end{equation}
This matches with the expression for the volume we derived earlier using the maximization technique. 

\section{Entropy in the Black Hole interior}

Having understood the geometry of a space-like maximal hypersurface inside a rotating BTZ black hole, we now try to gain some insight into the process of counting of the states in the phase space living in the black hole interior. Since the black hole evaporates through Hawking radiation, and shrinks, it is not in an equilibrium state. In general we cannot do a state counting in such a dynamical background. However the problem is not as critical as it may seem on its face. Christodoulou et al \cite{CT} have discussed that the maximal hypersurface which we derive using the geodesic method has a zero extrinsic curvature. We showed this earlier in section 3. And when we define a line element on such a hypersurface using the Gaussian coordinates, the surface does not change with the proper time. So we can indeed apply the techniques of quantum statistics to compute the entropy. Now we introduce a massless scalar field on the hypersurface $\Sigma$ and make an attempt to extract thermodynamimcal information. The scalar field will satisfy the Klein Gordon equation - $g_{\mu\nu}\partial^\mu \partial^\nu \phi = 0$. As we focus on the part of $\Sigma$ which corresponds to $\dot{r} = 0$, the line element would be $ds^2 = -f(r_{ss})dv^2 - Jdvd\theta + r_{ss}^2 d\theta^2$. Given this metric, 
\begin{equation}
g_{\mu\nu} = 
\Bigg( \begin{matrix}
-f(r_{ss})&-\frac{J}{2}\\  
\frac{-J}{2}&r_{ss}^2 
\end{matrix} \Bigg)
\end{equation}
Let $\lambda_1$ and $\lambda_2$ be the its eigenvalues. Then the Klein Gordan equation in the momentum space gives the following relation - 
\begin{equation}
E^2 = \lambda_1 p_v^2 + \lambda_2 p_\theta^2
\end{equation}
Now we count the number of states in the phase space which have energy less than E. This is given by 
\begin{equation}
g(E) = \frac{1}{(2\pi)^3}\int \sqrt{g}dvd\theta dp_v dp_\theta \end{equation}
Where we have put $\hbar = 1$. Substituting $P_v = \sqrt{\frac{E^2 - \lambda_2p_\theta^2}{\lambda_1}}$ and solving the momentum integral we get
\begin{eqnarray}
g(E) &=& \frac{1}{(2\pi)^3}\frac{\pi E^2}{2\sqrt{\lambda_1\lambda_2}}\int \sqrt{g} dv d\theta \\
&=& \frac{ E^2v}{8\pi}
\end{eqnarray}
The number of quantum states is proportional to the volume which is true for most systems. However in this case, we note that the volume of the hypersurface increases with the advanced time $v$. So there is always more space in the volume for the information to be stored even as the black hole evaporates through Hawking radiation and loses mass and the horizon shrinks with time. This has been discussed for other black hole families by \cite{CR,CT,YCO2}.
Next we compute the free energy for the scalar field at some inverse temperature $\beta$ as 
\begin{eqnarray} \nonumber
F(\beta) &=& \frac{1}{\beta} \int dg(E) ln (1 - e^{-\beta E})  \\ \nonumber
&=& -\frac{v}{8\pi} \int \frac{E^2 dE}{e^{\beta E} -1} \\ 
&=& -\frac{v}{8\pi\beta^3}  2\zeta(3) \\ \nonumber
\end{eqnarray}
We can then obtain the entropy as
\begin{eqnarray} \label{entropy}
S_\Sigma = \beta^2 \frac{\partial F}{\partial \beta} &=& \frac{3v}{4\pi\beta^2} \zeta(3) \\ \nonumber
\end{eqnarray}
Note that the entropy in the volume of a black hole is proportional to the advance time. This is certainly different from the entropy living on the black hole horizon $S_A$. To obtain a relation between $S_\Sigma$ and $S_A$, we need to restrict our parameter space. The horizon temperature of a BTZ black hole is given as
\begin{equation}\label{temp}
T_H = \frac{M}{2 \pi r_+}\sqrt{1 - \big(\frac{J}{M}\big)^2}
\end{equation}   
This temperature would in general not be equal to the temperature of the scalar field residing in the CR volume of the maximal hypersurface. We explore the regime when the black hole is close to the extremal limit when $M \rightarrow J$. Here it is important to understand the subtle point that the an exact extremal black hole demands a very different treatment. The CR volume for an extremal black hole would be exactly zero. However the volume keeps on increasing for a "near" extremal black hole and the instant at which the "near" extremal black hole turning into an "exact" extremal is not well understood. This point is extensively  discussed by Yen \cite{YCO}, \cite{YCO2}. Now in the near extremal limit, the loss of energy through the Hawking radiation is negligible. If we put $M = J + \delta$, where $\delta$ is a small number in eq. \ref{temp}, we get a relation $T_H \sim \sqrt{\delta}$. So in the near extremal limit, the horizon temperature is independent of the black hole mass as it approaches zero.  While the Hawking temperature varies continuously, if we start with a setup in which the scalar field is in thermal equilibrium with the horizon, it remains so. 
Under such a quasi-static conditions, we can find the relation between  the entropy $S_{\Sigma}$ and the horizon entropy $S_H$. For this we first identify the scalar field temperature $T = 1/\beta$ with the horizon temperature $T_H$. As the Hawking radiation is thermal in nature, the mass loss of the black hole per unit horizon area can be given using the Stefan-Boltzmann law for 2+1 dimensions - 
\begin{equation}
    \frac{1}{r_+}\frac{dM}{dv} = \sigma T_H^3
\end{equation}
Noting that $T_H \sim 1$ and $r_+ \sim \sqrt{M}$ in the extremal limit, we get $v \sim \sqrt{M}$. Plugging this expression in the eq. \ref{entropy}, we get
\begin{equation}
S_\Sigma \propto \frac{\sqrt{M}}{\beta^2}
\end{equation}
As discussed above, in the extremal limit $T_H$ is scale invariant and hence $\beta$ is also scale independent. So $S_\Sigma \propto \sqrt{M}$, which is an interesting result. We  find that the entropy of the quantum field in the maximal hypersurface is proportional to the horizon perimeter which is the boundary of that hypersurface. This result is similar to what was shown by Zhang et al \cite{BZ} for a Schwarzschild black hole and then also discussed extensively in Zhang \cite{BZ2}. As the entropy depends on the horizon area, our calculation exhibits compatibility with the laws of black hole thermodynamics.    

\section{Conclusion}
While the event horizon of a black hole and its area are well defined and an absolute quantity, computing the volume enclosed by the horizon can be a little tricky. \cite{CR} demonstrated a technique to compute the volume of a spherically symmetric spacelike hyper-surface which is bounded by the horizon. It was found that the maximum contribution to the volume comes from a particular  value of the radial coordinate $r$. In this work we extended the technique developed by \cite{CR} and derived the expression for the volume of a rotating black hole in 2+1 dimensions. Further we used the extremization technique developed by \cite{BZ} to compute the volume and the "steady state" radius. The volume grows linearly with the advance time and this provides a lot of room for information to be stored. We investigated this aspect for a near extremal 2+1 rotating black hole and computed the entropy of a scalar field living in the maximal hyper-surface. We  found that this entropy, $S_{\Sigma}$ turns out proportional to the horizon entropy $S_H$ and this conclusion is similar to the result obtained by \cite{BZ2} and \cite{MA2} for the case of BTZ black hole. Based on the techniques developed in this work, we can extend the same for the case of Kerr family of blackholes and develop the CR equations for the same (work in progress).

\section{Acknowledgement}
We all thank our institute BITS Pilani Hyderabad campus for providing the required infrastructure to carry out this work.




\end{document}